\documentclass[aps, reprint, twocolumn, floatfix,superscriptaddress,amsmath,amssymb,apsrev]{revtex4-1}
\usepackage[colorlinks,urlcolor=blue,linkcolor=blue,anchorcolor=blue,citecolor=blue]{hyperref}
\usepackage{graphicx}
\usepackage{epsfig}
\usepackage{wasysym}
\usepackage{soul}
\newcommand{\sumnear}{\mathop{\sum}_{\langle i j \rangle}}

\begin{document}
\title{Continuous easy-plane deconfined phase transition on the kagome  lattice}
\author{Xue-Feng Zhang}
\affiliation{Department of Physics, Chongqing University, Chongqing 401331, People's Republic of China}
\affiliation{Max-Planck- Institute for the Physics of Complex Systems, 01187 Dresden, Germany}
\author{Yin-Chen He}
\affiliation{Department of Physics, Harvard University, Cambridge, MA 02138, USA}
\author{Sebastian Eggert}
\affiliation{Physics Department and Research Center OPTIMAS, University of Kaiserslautern, 67663 Kaiserslautern, Germany}
\author{Roderich Moessner}
\affiliation{Max-Planck- Institute for the Physics of Complex Systems, 01187 Dresden, Germany}
\author{Frank Pollmann}
\affiliation{Department of Physics, Technical University of Munich, 85748 Garching, Germany}
\affiliation{Max-Planck- Institute for the Physics of Complex Systems, 01187 Dresden, Germany}
\date{\today}

\begin{abstract}
We use large scale quantum Monte-Carlo simulations to study an extended Hubbard model of hardcore bosons on the kagome lattice. In the limit of strong nearest-neighbor interactions at $1/3$ filling, the interplay between frustration and quantum fluctuations leads to a valence bond solid ground state. The system undergoes a quantum phase transition to a superfluid phase as the interaction strength is decreased. It is still under debate whether the transition is weakly first order or represents an unconventional continuous phase transition. We present a theory in terms of an easy-plane NCCP$^1$ gauge theory describing the phase transition at 1/3 filling. Utilizing large scale quantum Monte-Carlo simulations with parallel tempering in the canonical ensemble up to 15552 spins, we provide evidence that the phase transition is continuous at exactly $1/3$ filling. A careful finite size scaling analysis reveals an unconventional scaling behavior hinting at deconfined quantum criticality. 

\end{abstract}


\maketitle

\textbf{Introduction.} Understanding universal and non-universal properties of quantum phase transitions in strongly correlated systems is a key topic in modern physics \cite{sachdev2007quantum}.
In many cases, quantum phase transitions can be described by Landau's theory of spontaneous symmetry breaking just like classical ones.
On the other hand, there appear to exist exotic quantum phase transitions beyond the Landau's paradigm such as continuous deconfined phase transitions (DCPs) between phases with different, incompatible symmetry breakings.
A well known example is the transition between a N\'{e}el state and a valence bond solid (VBS) \cite{deccp,deccplong}.

Contrary to conventional phase transitions, a deconfined phase transition  exhibits fractionalized quasiparticles that couple to emergent gauge fields \cite{deccp,deccplong}.
Deconfined phase transition are generically described by strongly interacting gauge theories.
One example is the non-compact  $\mathbb{CP}^1$ (NCCP$^1$) model with a bosonic $\mathbb{CP}^1$ field $z_\alpha$ (describing spinons with ${\mathrm{SU}(2)}$ flavors $\alpha = 1,2$), which couples to a non-compact $U(1)$ gauge field $a_\mu$. 
Depending on the symmetries of the field $z_\alpha$, the  NCCP$^1$ models are divided into $\mathrm{SU}(2)$-NCCP$^1$ and the easy-plane-NCCP$^1$, which describe the N\'{e}el to VBS transition in  ${\mathrm{SU}(2)}$ or XY magnets \cite{deccp,deccplong,lesikav04}, respectively.

The concept of DCPs leads to several interesting questions:
First, to which extent  do these emergent gauge fields and fractionalized excitations appear at critical points in concrete model systems?
Second, what is the fate of the NCCP$^1$ model in the infrared (IR) limit? 
Recent progress in the understanding of dualities of gauge theories has brought new perspectives to deconfined phase transitions~\cite{wangsenthil15b,MaxAshvin15,dualdrMAM,qeddual,lesikav04,tsmpaf06,seiberg1,karchtong,WangDCPdual,dualdrMAM2}.
It has been conjectured that the bosonic easy-plane NCCP$^1$ theory is dual to a widely studied fermionic $N_f=2$ QED$_3$ theory~\cite{tsmpaf06,qeddual,seiberg1,karchtong,WangDCPdual,dualdrMAM2}.
Significant effort has been put into the investigation of the IR fate of QED$_3$, but it remains an open issue after several decades of study~\cite{QEDCSB1,QED3CFT,kogut2,GroverFT,QED3RG,QEDQMC1,QEDQMC2,qedcft}.
Studying concrete realizations of DCPs helps to deepen the understanding of this long-standing problem.

Numerical work~\cite{SandvikJQ,melkokaulfan,lousandvikkawashima,Banerjeeetal,Sandviklogs,sandvik2parameter,dcphex2013,Kawashimadeconfinedcriticality,Jiangetal,deconfinedcriticalityflowJQ,DCPscalingviolations,emergentso5,MotrunichVishwanath2,kuklovetalDCPSU(2),CharrierAletPujol, Chenetal,Aletextendeddimer,powellmonopole,easydcpprl2006,easydcpprb2007,easydcpprb2016,easydcparxiv2016,ZYM} has studied both the $\mathrm{SU}(2)$ and easy-plane DCPs---most of them focused on the J-Q model ~\cite{SandvikJQ,melkokaulfan,lousandvikkawashima,Banerjeeetal,Sandviklogs,sandvik2parameter,dcphex2013,Kawashimadeconfinedcriticality,Jiangetal,deconfinedcriticalityflowJQ,easydcpprb2016,easydcparxiv2016} and classical loop models~\cite{powellmonopole,emergentso5,DCPscalingviolations}.
It is still controversially discussed if $\mathrm{SU}(2)$ DCPs are continuous~\cite{kuklovetalDCPSU(2)}, and an emergent $\mathrm{SO}(5)$ symmetry is observed~\cite{emergentso5} between N\'{e}el and VBS phases.
The easy-plane case on the other hand appears to be a first order transition in all previous numerical studies on various candidate model systems \cite{easydcpprl2006,easydcpprb2007,easydcpprb2016,easydcparxiv2016,kukloveasyplane}.
The question arises whether the easy-plane-NCCP$^1$ is intrinsically first order or if it is specific to the models that have been studied so far.

In this paper, we provide numerical evidence for the existence of a continuous easy-plane-DCP using large scale quantum Monte-Carlo simulations, our results are in agreement with a parallel work \cite{ZYM}.
Specifically, we study an extended Hubbard model of hardcore bosons on the kagome lattice,
\begin{eqnarray}\label{eq:Ham}
H&=&-t\sumnear(b_{i}^{\dag}b_{j}+b_{j}^{\dag}b_{i})+V\sumnear
n_{i}n_{j},\label{eq:ham}
\end{eqnarray}
at $1/3$ filling with $t,V>0$.
The system is known to form a VBS ground state in the limit $V\gg t$ and a superfluid for $V\ll t$; where both phases are separated by a quantum phase transition \cite{kagome1,kagome2,kagome3,RGmelko}.
We first discuss the easy-plane NCCP$^1$ theory~\cite{deccp,deccplong} that describes the superfluid-VBS transition.
In particular, we highlight the difference between our system and other systems hosting DCPs (e.g. the J-Q model).
By using large scale quantum Monte Carlo methods with parallel tempering (QMC-PT) in the canonical ensemble, we find that the phase transition between VBS to the superfluid is anomalously  continuous at exactly $1/3$ filling.
Several hallmarks of DCP are found: 
(i) At the critical point, the superfluid density decays slower than at regular continuous phase transitions. Comparing with different scenarios~\cite{sandvik2parameter,dcphex2013,DCPscalingviolations}, we adopt logarithmic corrections to fit this drift. 
(ii) A direct analysis of  two point correlations reveals that the anomalous critical exponent $\eta\approx0.3$ is relatively large.
(iii) We identify a lattice operator for a conserved charge (i.e. the spinon density) of NCCP$^1$, and numerically show that its scaling dimension is close to two, as expected for a 2+1D conformal field theory (CFT) \cite{francesco2012conformal}.
(iv) An emergent $U(1)$ symmetry is identified at the critical point.
\begin{figure}[t]
\includegraphics[width=0.45\textwidth]{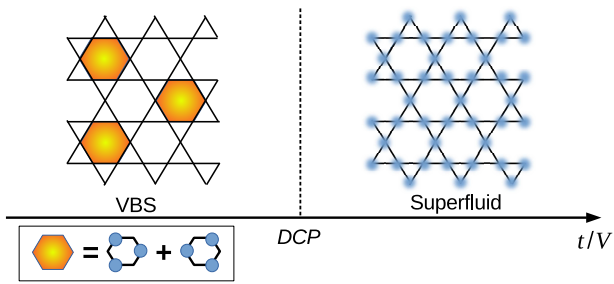}
\caption{Phase diagram of Hamiltonian (\ref{eq:ham}) at $1/3$ filling with a DCP separating a VBS and a superfluid phase. In the VBS phase resonant process (colored hexagon) spontaneously break translation symmetry. In contrast, in the superfluid phase the bosons condense and spontaneously break the $U(1)$ symmetry \label{fig1}.}
\end{figure}

\textbf{Effective theory and phases}. Similar to the much studied N\'{e}el-VBS transition in antiferromagnets~\cite{deccp,deccplong}, the superfluid-VBS transition in our system is also described by the   NCCP$^1$ theory,
\begin{equation}
\label{nccp1easy} {\cal L} = \sum_{\alpha = 1}^2 \left[ |(\partial_\mu -ia_\mu) z_\alpha|^2
+s |z_\alpha|^2+u|z_\alpha|^4\right]+v|z_1|^2|z_2|^2+\cdots,
\end{equation}
where the $z_{\alpha=1,2}$ are bosonic ($\mathbb{CP}^1$) fields (or spinon operators) carrying half the charge of the physical bosons, and they are coupled to an emergent dynamical $U(1)$ gauge field $a_\mu$.
The mass term $s|z_\alpha|^2$ with $s\approx V/t-7$ controls the phases:
(i) if $z_\alpha$ condenses, a superfluid phase is formed; 
(ii) if $z_\alpha$ is gapped, a VBS state forms due to the proliferation of monopoles of the gauge field $a_\mu$ \cite{ReSaSUN}
(iii) the case of $z_\alpha$ being gapless corresponds to the critical point.
The quartic terms with $u\approx O(t^3)$ and $v\approx O(t^3)$ control the putative IR fixed point to which the theory flows under renormalization.
When $v=2u$, there is a $\mathrm{SU}(2)$ symmetry between $z_1$ and $z_2$, and the theory is called $\mathrm{SU}(2)$ NCCP$^1$. 
Usually the $\mathrm{SU}(2)$ symmetry will be manifest  as a global $\mathrm{SU}(2)$ symmetry of the spin system.
On the other hand if $v<2u$, the theory flows to the easy-plane NCCP$^1$ fixed point where the $\mathrm{SU}(2)$ symmetry is broken.
Our hardcore boson model Eq.~\eqref{eq:Ham} naturally falls into the easy-plane NCCP$^1$ class.
The same field theory also describes the N\'{e}el-VBS transition in other related spin models (e.g. J-Q model), and the hard-core boson model we are studying can be exactly mapped to a spin-1/2 model by $b^\dag \rightarrow S^+$, $n=b^\dag b \rightarrow S^z+1/2$.

In our system, however, the relation between the continuous field operator and the lattice operators is very different from the usual DCP in spin models.
In  usual spin models (e.g. the J-Q model), one would have $S^+\sim z_1^* z_2$, $S^z\sim |z_1|^2-|z_2|^2$. 
In our case, the relations are
\begin{align}
b^\dag_i & \sim z_1^* z_2,\quad   n_i=b^\dag_i b_i \sim E_i + \textrm{Re}(e^{i\theta_i} \mathcal M_a), \label{fieldtospin1} \\
\sum_{i\in \bigtriangleup} n_i  & \sim z_1^* \partial_t z_1,\quad \sum_{i\in \bigtriangledown} n_i \sim z_2^* \partial_t z_2. \label{fieldtospin2}
\end{align}
Here $E_i$ represents the electric fields of the dynamical gauge field $a_\mu$, $\mathcal M_a$ is the monopole operator while $\theta_i$ is a phase factor ($=0, \pm2\pi/3$) depending on the sublattice index.
$\sum_{i\in \bigtriangleup,\bigtriangledown } n_i $ refers to the summation of the density of three sites in the up or down triangles of the kagome lattice. 
The difference originates from the different fractionalization schemes of the spin operator $\vec S$ into the $\mathbb{CP}^1$ (spinon) field $\mathbf z = (z_1, z_2)$.  
Usually at DCPs, the spin operator is fractionalized via the $\mathbb{CP}^1$ representation  $\vec S_i = \mathbf z_i^* \vec \sigma \mathbf z_i^T$~\cite{deccp,deccplong}, and such a spinon operator is argued to capture the low energy physics.
In contrast, our kagome  model can be faithfully mapped onto a lattice gauge model defined on the medial honeycomb lattice \cite{Nikolic2005, YCH15_2}, in which spinons ($z_{1,2}$) live on honeycomb sites (i.e. center of kagome triangles) and $\mathrm{U}(1)$ gauge fields live on the honeycomb links. 
Then we can straightforwardly take the continuum limit of the lattice gauge model, which precisely gives the easy-plane NCCP$^1$ theory.
%

The relations in Eq.~\eqref{fieldtospin1}-\eqref{fieldtospin2} call for a slightly different way of extracting critical exponents. 
Specifically, the anomalous dimension $\eta_{\textrm{VBS}}$ of the VBS order parameter should be extracted from the density operator $n_i$, instead of the dimer operator in the J-Q model.
The operators $s_{\bigtriangleup,\bigtriangledown}=\sum_{i \in \bigtriangleup, \bigtriangledown} n_i$, correspond to conserved charges of the gauge theory, $z^*_\alpha \partial_t z_\alpha$. 
For any $2+1$D CFT, such a conserved charge will always have scaling dimension two \cite{francesco2012conformal}, providing an additional numerical check.

\textbf{Numerical results.} We use a stochastic cluster series expansion with parallel tempering \cite{sse1,sse2,sse3,sse4} and adopt periodic boundary conditions with $L_x=L_y$. 
To reach the ground state, we use half million steps of thermalization before producing two million samples for measuring and consider temperatures down to $\beta V/L=25/3$ ($\beta=1/T$).
We identify the diagonal order in the VBS phase using the structure factor  $S(\mathbf{Q})=\langle |n(\mathbf{Q,\tau})|^2\rangle=\langle |\sum_{k=1}^N n_{k,\tau} e^{\mathbf{iQ}\cdot\mathbf{r}_k}|^2\rangle/N^2$ at $\mathbf{Q}=(4\pi/3,0)$ where $N$ is the number of sites.
For the superfluid phase, we consider the superfluid density $\rho_s=\langle W^2\rangle/\beta t$ where $W$ is the winding number \cite{winding} and also the condensate fraction $\rho_0=\langle \sum_{i,j}b_i^{\dagger}b_j\rangle/N^2$ to characterize long range off-diagonal correlations.
\begin{figure}[t]
\includegraphics[width=0.49\textwidth]{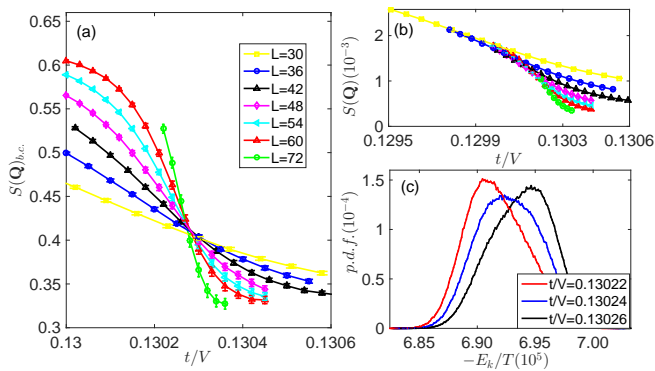}%
\caption{The structure factor (b) and its Binder cumulant (a) vs.  $t/V$ at $1/3$ filling and $\beta V/L=25/3$ with different $L$. (c) The probability density function (p.d.f) of  kinetic energy at $L=72$ and $\beta V=600$ near the critical point.\label{fig2}}
\end{figure}

It turns out that a continuous phase transition only occurs at exactly $1/3$ filling where the system has particle-hole symmetry.
In previous studies \cite{kagome1,kagome2}, a grand canonical ensemble was used for the QMC simulations which made it difficult to fine tune to exactly $1/3$ filling. 
Here, we restrict our simulations to the canonical ensemble by tuning the chemical potential to minimize the deviation from $1/3$ filling during the loop-update and then only accept samples with exactly $1/3$ filling. 
From Fig.~\ref{fig2}, we find (i) the structure factor $S(\mathbf{Q})$ does not show any discontinuity for sizes up to a linear dimension of $L=72$ ($N=15552$ spins); (ii) its Binder cumulant $S(\mathbf{Q})_{b.c.}=1-\frac{\langle S(\mathbf{Q})^2 \rangle}{3\langle S(\mathbf{Q}) \rangle^2}$ is always positive and crosses at approximately same point $t_c/V\approx 0.1303$; (iii) at variance from Ref.\cite{kagome1}, at larger size $L=72$ near the critical point, we do not find any double peak structure in the probability distribution of kinetic energy. Since the parameter $t/V=0.1283$ in Ref. \cite{kagome1} 
is actually far from $t_c/V$, it reflects the weakly first order phase transition at the upper/lower boundary of the lobe, but not at the tip. These three findings strongly support a continuous phase transition up to system size $L=72$.

Next we perform finite size scaling (FSS) for different variables to extract the critical behavior.
For a continuous phase transition, the scaling function takes the form:
\begin{eqnarray}
A(L,\delta)&=&L^{-\kappa} f(\delta L^{1/{\nu}}), \label{fss}
\end{eqnarray}
where $\nu$ and $\kappa$ are related to the universality class of the phase transition, and $\delta=t/V-t_c/V$. Because the form of the scaling function $f$ is not known, we choose the method of Kawashima and Ito proposed to do the data collapse \cite{KI1,KI2}.

\begin{figure}[ht]
\includegraphics[width=0.49\textwidth]{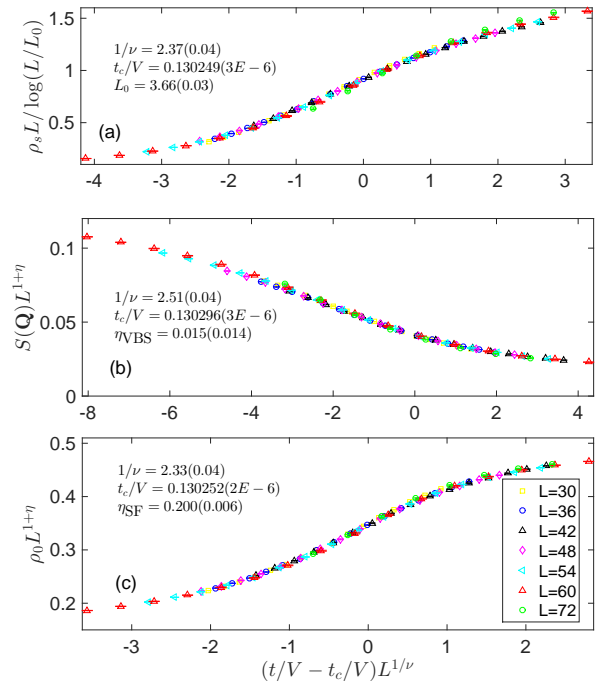}%
\caption{Data collapse of (a) superfluid density with logarithmic correction, (b) structure factor and (c) condensate fraction at $1/3$ filling and $\beta V/L=25/3$. \label{fig3}}
\end{figure}

An anomalous behavior of FSS of physical quantities has been observed in all  previous numerical works on DCP (see examples~\cite{SandvikJQ,melkokaulfan,Banerjeeetal,Sandviklogs,sandvik2parameter,dcphex2013,DCPscalingviolations,emergentso5,MotrunichVishwanath2}).
It has been  suggested that these anomalous scaling behaviors arise due to finite size effects of dangerously irrelevant operators \cite{DCPscalingviolations}.
For example, the superfluid density $\rho_s$ shows a drift~\cite{sandvik2parameter,dcphex2013,DCPscalingviolations} compared to the scaling of the conventional phase transition,   $\rho_s(L,\delta)=L^{-1}f(\delta L^{1/\nu})$.
To resolve the drift, two schemes have been proposed: (i) logarithmic corrections (LCs)  $\rho_s(L,\delta)=L^{-1}\log(L/L_0)f(\delta L^{1/\nu})$~\cite{Sandviklogs,DCPscalingviolations} and (ii) two-length scales $\rho_s(L,\delta)=L^{-\nu/\nu'}f(\delta L^{1/\nu})$~\cite{sandvik2parameter}.
In our work we use the LCs and find a good data collapse with $1/\nu=2.37(0.04)$, as shown in Fig. \ref {fig3}a. Using two-length scales also gives a reasonably good collapse \cite{sup}.

Scaling violations are also observed in the diagonal structure factor $S(\mathbf{Q})$ and condensate fraction $\rho_0$, whose FSS has previously been used to extract the anomalous dimension $\eta_{\text{VBS}}$ and $\eta_{\text{SF}}$.
The anomalous dimensions $\eta_{\mathrm{VBS}}=0.015(0.014)$ and  $\eta_{\mathrm{SF}}=0.200(0.006)$ extracted from $S(\mathbf{Q})$ and $\rho_0$, respectively, strongly deviate from each other (shown in Fig.~\ref{fig3}b and  Fig.~\ref{fig3}c). This is not expected for the easy-plane NCCP$^1$ theory as both anomalous exponents are the same due to self-duality \cite{lesikav04}. Previous
studies~\cite{monopole3,DCPscalingviolations} find a large drift of the critical exponent and anomalous dimension due to the $1/L^2$-correction~\cite{footnote}. Therefore, a simple data collapse does not give good results, but by using the two-point correlator a size-dependent anomalous dimension can be extracted, which shows a systematic convergence to the thermodynamic limit as discussed in the following and in the Supplemental Material~\cite{sup}.

\begin{figure}[t]
	\includegraphics[width=0.49\textwidth]{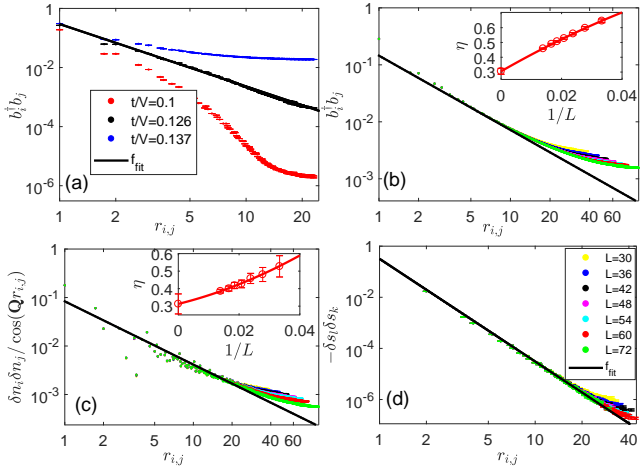}%
	\caption{
		(a) Off-diagonal correlation function vs. distance at $1/3$ filling for $L=18$, $\beta V=300$, and different $t/V$. 
		(b) Off-diagonal and (c) diagonal correlation function vs. distance for different system sizes near the critical point at $1/3$ filling, $t/V=0.1303$ and $\beta V/L=25/3$. Inset: FSS of the anomalous exponents from fitting to $C_l(\delta,r)$. The extracted critical exponents are $\eta_{\text{SF}}=0.305(0.020)$ and  $\eta_{\text{VBS}}=0.313(0.057)$. 
		(d)  Spinon correlation function vs distance at $1/3$ filling,  near the critical point  $t/V=0.1303$ and $\beta V/L=25/3$ for different system sizes. The black line is a fit to a power law decay $r^{-4.02(0.18)}$.\label{fig4}}
\end{figure}

We find two different scaling behaviors as we approach the continuous quantum phase transition from the two neighboring phases:
In the disordered phase  $C_s(\delta,r)=ar^{-1-\eta}\exp(-r/\xi(\delta))$ with  the correlation length $\xi(\delta)\propto\delta^{-\nu}$, while approaching the critical point from the ordered phase, $C_l(\delta,r)=ar^{-1-\eta}+b(\delta)$ ($b(0)=0$) \cite{sachdev2007quantum}. 
At the critical point we then expect a power law decay  $C_l=ar^{-1-\eta}$. 
As shown in Fig. \ref{fig4}a, the off-diagonal correlation function $\langle b_i^{\dag} b_j \rangle$ decays very fast in the VBS phase ($t/V=0.1$) which hints at an exponential behavior, while it decays slowly to a constant in the superfluid phase ($t/V=0.137$). Near the critical point ($t/V=0.126$), it shows a clear power law behavior.  
To approach the thermodynamic limit (TDL), we calculate the correlation function near the critical point ($t/V=0.1303$), and perform a FSS analysis on the exponent.  
As shown in Fig.~\ref{fig4}b, we identify a power law decay with increasing system size. 
The inset of Fig.~\ref{fig4}b shows strong finite size effects of the anomalous exponent, and these size effects can extremely depress the exponent obtained from the data collapse of the condensate fraction~\cite{sup}. With second order polynomial fitting, we get $\eta_{\text{SF}}=0.305(0.020)$ in the TDL.
Fig.~\ref{fig4}c shows the density correlation function for different parameters. 
Contrary to the off-diagonal correlations, the density correlation functions have a density modulation due to translational symmetry breaking in the VBS. We thus subtract its mean value and divide by the density modulation $\cos(\textbf{Q}r_{i,j})$. 
While the correlations show strong fluctuations at short distance, a smooth power law decay emerges at long distances and we thus neglect the first ten points for the fitting.
Comparing to the off-diagonal correlations, the error is larger and the anomalous exponent in the TDL is $\eta_{\text{VBS}}=0.313(0.057)$.

From Eq.~\eqref{fieldtospin2} we identify the spinon density $s_{\bigtriangleup,\bigtriangledown}=\sum_{i \in \bigtriangleup, \bigtriangledown} n_i$  as a conserved charge.
Such conserved charge should have scaling dimension $\Delta=2$ for any $2+1$D CFT.
As shown in Fig. \ref{fig4}d, the corresponding correlation function shows a fast power law decay and rather small finite size effects. 
The extracted exponent is $-4.02(0.18)$, which is relatively close to $-2\Delta$, strongly supporting the scenario that the easy-plane-NCCP$^1$ is a CFT.

\begin{figure}[t]
\includegraphics[width=0.5\textwidth]{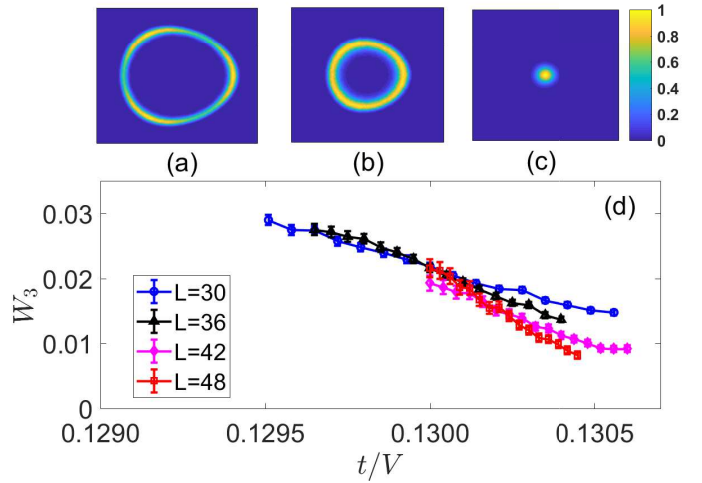}%
\caption{Histogram of [$\Xi(\mathbf{Q}_h)_x$, $\Xi(\mathbf{Q}_h)_y$], in (a) VBS phase ($t/V=0.124$), (b) near the critical point ($t/V=0.129$) and (c) superfluid phase ($t/V=0.131$) at $1/3$ filling, $L=30$ and $\beta V=500$. (d) The $Z_3$ anisotropy parameter $W_3$ at $1/3$ filling and $\beta V/L=25/3$ for different system sizes.  \label{fig5}}
\end{figure}

A hallmark of DCPs are emergent symmetries \cite{deccplong, emergentso5, WangDCPdual}.
For example, at the critical point, the lattice $Z_3$ rotation symmetry will be enlarged to a continuous $U(1)$ rotation symmetry.
To check this, we consider the resonant valence bond order parameter $\Xi(\mathbf{Q}_h)=\sum_{k=1}^{N_h} \Xi_k e^{\mathbf{iQ_h}\cdot\mathbf{r}_k}/\sqrt{N_h}$ with $\mathbf{Q_h}=(2\pi/3,0)$ where $\Xi_k$ is the density operator of a hexagon of a resonant configuration, and $N_h$ is the number of hexagons. 
In the VBS phase, the $Z_3$ degeneracy implies the phases of $\Xi(\mathbf{Q}_h)$ are $0$, $2\pi/3$ and $4\pi/3$. We define $\Xi(\mathbf{Q}_h)_x$ ($\Xi(\mathbf{Q}_h)_y$) as the real (imaginary) part $\text{Re}[\Xi(\mathbf{Q}_h)]$ ($\text{Im}[\Xi(\mathbf{Q}_h)]$) of $\Xi(\mathbf{Q}_h)$. 
From the histogram shown in Fig.~\ref{fig5} we find that it has three peaks in the VBS phase, indicating $Z_3$ symmetry, which shrinks to one point in the superfluid phase, reflecting no solid order. 
Near the critical point, the distribution approaches a uniform circle which reveals emerging $U(1)$ symmetry. 
In order to quantitatively check this, we introduce a $Z_3$ anisotropy parameter $W_3=\langle\cos(3\arg(\Xi(\mathbf{Q}_h)))\rangle$.
Fig.~\ref{fig5}d shows this quantity to increase in the VBS phase and to vanish in the superfluid phase with increasing system size.

\textbf{Conclusions and discussions}. We have studied the easy-plane deconfined phase transition of a hard-core Bose-Hubbard model using QMC. Finite size simulations of clusters up to $L=72$ indicate an anomalous critical point separating the VBS and superfluid phase. We estimate the critical
point is at $t_c/V\approx0.1303$. Following the approach in Ref.\cite{DCPscalingviolations}, we extract the anomalous exponents  $\eta_{\text{SF}}\approx0.305$ and $\eta_{\text{VBS}}\approx0.313$ from the two-point correlation functions.
In addition, we identify a lattice operator for the conserved charge of NCCP$^1$, and we numerically show its scaling dimension is $\Delta$$\approx$$2$. At last the emergent U(1)
rotation symmetry is found at the critical point. 

Comparing with another easy-plane NCCP$^1$ model~\cite{kukloveasyplane}, our model can be viewed as a different way to regularize an  easy-plane NCCP$^1$ continuous field theory on a discrete lattice. For example, in our system there is only one $U(1)$ global symmetry, while in the paper by Kuklov \textit{et. al.} there are two $U(1)$ global symmetries. This difference leads to the winding number of two type spinons in our model are equal $W_-=W_{\bigtriangleup}-W_{\bigtriangledown}=0$ which means no super-counter fluid phase exists. Such difference may also strongly change the type of phase transition. 
Altogether, our results strongly support the presence of easy-plane deconfined criticality. Similar to previous works, our data shows some scaling violation that require further studies.

We are thankful for useful discussions with  Adam Nahum, Arnab Sen, Yuan Wan, G. J. Sreejith, Wenan Guo, Stefan Wessel, and Chong Wang. This work was
supported in parts by the German Research Foundation (DFG) via the
Collaborative Research Centers SFB/TR49, SFB/TR173, SFB/TR185, SFB 1143 and Research Unit FOR 1807 through grants no. PO 1370/2-1. 
The authors gratefully acknowledge the computing time granted by the John von Neumann Institute for Computing (NIC) on the supercomputer JURECA at
J\"{u}lich Supercomputing Centre (JSC), by the Allianz f\"{u}r
Hochleistungsrechnen Rheinland-Pfalz (AHRP) and by the Max-Planck Computing and Data Facility (MPCDF). 
YCH is supported by a postdoctoral fellowship from the Gordon and Betty Moore Foundation, under the EPiQS initiative, GBMF4306, at Harvard University. 

\bibliography{DCP}

\begin{thebibliography}{65}
\bibitem [{foo()}]{footnote}%
  \BibitemOpen
  \href@noop {} {\ }\bibinfo {note} {{We try to include this correction term
  $aL^{-2}$\cite{Deng_fit}, but the best fitting is at its prefactor
  $a=0$}.}\BibitemShut {Stop}%
\end{thebibliography}
\newpage
\centering{\appendix{\textbf{Supplementary Material}}}
\section{Finite size scaling}
At the critical point of a second order phase transition, the two-point correlator follows a power lay decay $C_L(r)=a(L)r^{-1-\eta(L)}$. In large systems, the structure factor or condensate fraction is proportional to its integration per site:
\begin{eqnarray}
\int_1^L C_L(r)rdr/L^2=\frac{a(L)}{1-\eta(L)}(L^{-1-\eta(L)}-L^{-2}). \label{fss}
\end{eqnarray}
\begin{figure}[ht]
	\includegraphics[width=0.48\textwidth]{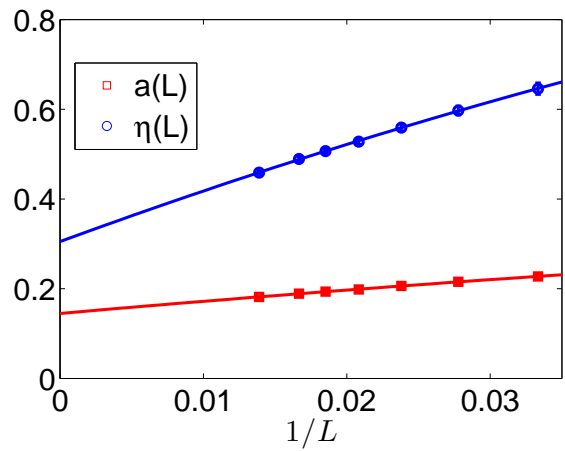}%
	\caption{The coefficients $a(L)$ and $\eta(L)$ got from off-diagonal two-point correlator in different sizes, by using power fitting $C_L(r)$ at $1/3$ filling, $t/V=0.1303$ and $\beta V/L=25/3$. \label{sxy_eta}}
\end{figure}

As mentioned in main text, we can use $C_L(r)$ to fit the two-point correlator function. For the off-diagonal correlator $b_i^{\dagger}b_j$ in Fig.\ref{sxy_eta}, we can find the prefactor $a(L)$ changes less than $\eta(L)$. Then, we use a second order polynomial function $\eta_{f}(L)$ to fit $\eta(L)$:
\begin{eqnarray}
\eta_{f}(L)=\eta_0+\eta_1/L+\eta_2/L^2,
\end{eqnarray}
which gives $\eta_0=0.305(0.020)$. We also considered higher order polynomial fitting, but the third order $\eta_0=0.299(0.113)$ has much less accuracy and higher orders are even overfitted.

In order to check how finite size effects change the anomalous exponent, we substitute $\eta(L)$ with fitting function $\eta_{f}$ in Eqn.(\ref{fss}) and get
\begin{eqnarray}
f_s\equiv\frac{a_{f}(L)}{1-\eta_{f}(L)}(L^{-1-\eta_{f}(L)}-L^{-2}),
\end{eqnarray}
where $a_f(L)$ is a second order polynomial fitting function for $a(L)$, and condensate fraction should be proportional to $f_s$. Then, we analyze the finite size effect 
of $a_f(L)$, $L^{-2}$ and $\eta_f(L)$ separately by defining:
\begin{eqnarray}
f_a&\equiv&\frac{a_0}{1-\eta_{f}(L)}(L^{-1-\eta_{f}(L)}-L^{-2}),\\
f_2&\equiv&\frac{a_0}{1-\eta_{f}(L)}(L^{-1-\eta_{f}(L)}),\\
f_{\infty}&\equiv&\frac{a_0}{1-\eta_0}(L^{-1-\eta_0}).
\end{eqnarray}
\begin{figure}[t]
	\includegraphics[width=0.48\textwidth]{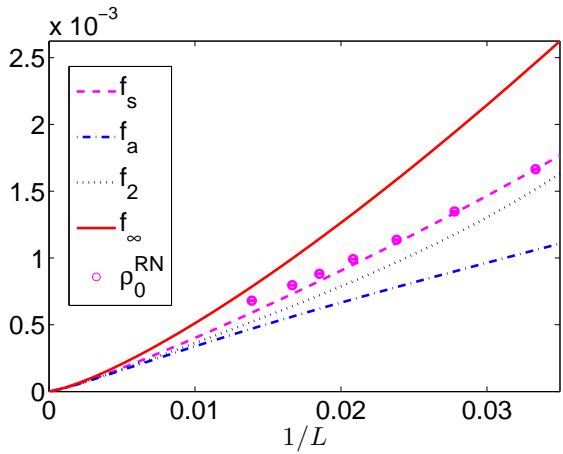}%
	\caption{Comparing condensate fraction with different functions which neglects various size effect terms\label{sxy_fss}}
\end{figure}

\begin{figure}[t]
	\includegraphics[width=0.48\textwidth]{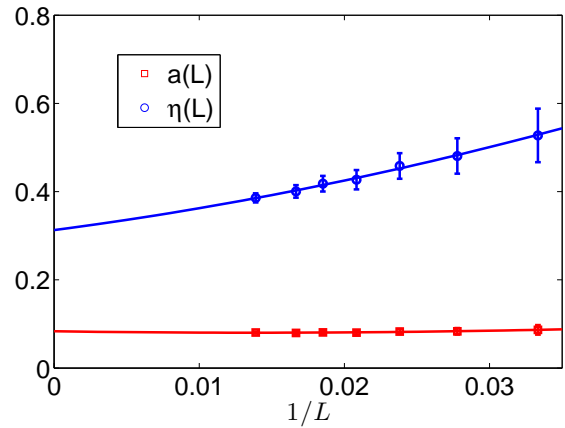}%
	\caption{The coefficients $a(L)$ and $\eta(L)$ got from density two-point correlator in different sizes, by using power fitting $C_L(r)$ at $1/3$ filling, $t/V=0.1303$ and $\beta V/L=25/3$. \label{szz_eta}}
\end{figure}

\begin{figure}[t]
	\includegraphics[width=0.48\textwidth]{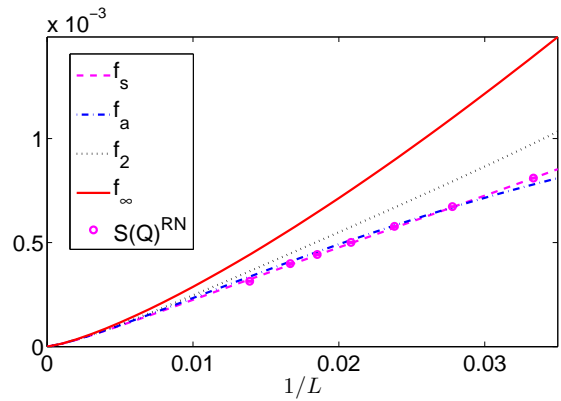}%
	\caption{Comparing structure factor with different functions which neglects various size effect terms. \label{szz_fss}}
\end{figure}

As shown in Fig.\ref{sxy_fss}, $f_s$ matches well with $\rho_0$ after rescaling its magnitude ($\rho_0^{RN}=C \rho_0$). Both $a_f$ and $L^{-2}$ terms can change the shape of the curve. However, the finite size effect of $\eta_{f}$ bends the curve which explains why the anomalous critical exponent obtained from a data collapse of the condensate fraction deviates strongly. 
A similar phenomenon also happens for the density correlator and structure factor show in Fig.\ref{szz_eta} and Fig.\ref{szz_fss}. Therefore the size independent critical exponents got directly from structure factor and condensate fraction are less convincing.

\begin{figure}[ht]
	\includegraphics[width=0.49\textwidth]{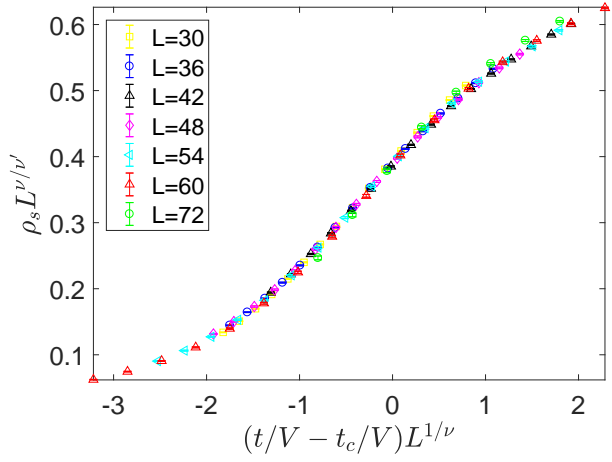}%
	\caption{Data collapse of superfluid density with two-length scales scenario at $1/3$ filling and $\beta V/L=25/3$. The critical point is $t_c/V=0.130263(0.000003)$, and the critical exponents are $1/\nu=2.299(0.041)$ and $\nu/\nu'=0.524(0.014)$. \label{rhos}}
\end{figure}

In addition, we also consider the data collapse of superfluid density with two-length scale scenario. As shown in Fig.\ref{rhos}, it is reasonable as good as LC scenario, so we can not conclude which one is better.

\section{numerical flowgram}
The numerical flowgram method was introduced by A.B. Kuklov, {\it et. al.}~\cite{kukloveasyplane,kuklovetalDCPSU(2),kuk} to study the DCPs. They obtain the finite size critical point $t_c(L)/V$ from the condition that the ratio of probabilities of having zero and non-zero winding numbers is some fixed number of the order of unity. If the transition is continuous, they claim the winding number at this critical point will approach a universal value when enlarging the system, otherwise, it will linearly scale with system size $L$.  

\begin{figure}[t]
	\includegraphics[width=0.48\textwidth]{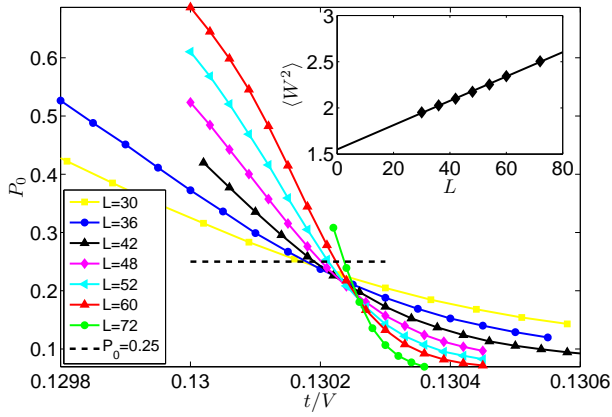}%
	\caption{The possibility of zero winding number vs $t/V$ for different sizes at $1/3$ filling and $\beta V/L=25/3$. Inset: The winding numbers for different size at finite size critical points where $P_0=0.25$ (dash line in main panel). \label{p}}
\end{figure}

We also implement numerical flowgram method for our case. We fix the probability of zero winding numbers $P_0$ equal to $0.25$ which marked in Fig. \ref{p}. We determine the size-dependent hopping values $t_{P_0}(L)$ for this probability. The corresponding winding numbers for those parameters can then be analyzed as a function of length $L$. As shown in inset of Fig.\ref{p}, the winding numbers don't flow to a universal value, but seem to linearly depend on the system size. Similar behaviour is also found in the J-Q model~\cite{kuk}, and it may be directly related to the drift of the superfluid density or weakly first order phase transition. 
\end{document}